\pdfoutput=1
\documentclass[journal]{IEEEtran}
\usepackage{amssymb}
\usepackage{amsmath}
\usepackage{graphicx}
\usepackage{cite}
\usepackage[T1]{fontenc}
\usepackage{txfonts}
\usepackage{bbm}
\usepackage[caption=false,font=footnotesize]{subfig}
\usepackage{flushend}
\begin{document}

\title{Feedback-Free FDD Cell-Free Massive MIMO With Sparse AP Activation and Dual-Mode Precoding}

\author{Wei~Jiang,~\IEEEmembership{Senior~Member,~IEEE,}~and~Hans~Dieter~Schotten\thanks{Manuscript received November 19, 2026. (\textit{Corresponding author: Wei Jiang.})}
\thanks{W. Jiang and H. D. Schotten are with the German Research Center for Artificial Intelligence (DFKI), 67663 Kaiserslautern, Germany, and also with the Department of Electrical and Computer Engineering, University of Kaiserslautern (RPTU), Germany. (e-mail: wei.jiang@dfki.de, schotten@eit.uni-kl.de).}
}
\markboth{Submitted to IEEE Transactions on Vehicular Technology}{Jiang \MakeLowercase{\textit{and}} Schotten: Feedback-Free FDD Cell-Free Massive MIMO}
\maketitle

\begin{abstract}
This correspondence enables practical frequency-division duplex (FDD) operation in cell-free massive MIMO via two mechanisms. Sparse access-point (AP) activation reduces downlink pilot overhead to enable coherent detection, while angle-reciprocity precoding steers beams using only uplink-estimated angles, eliminating channel feedback entirely. A load-adaptive dual-mode optimizer further enhances performance: closed-form signal-to-leakage-and-noise-ratio precoding for high-load rate boosting, or semidefinite relaxation for low-load power minimization. Simulations verify its superiority over existing feedback-free FDD solutions in spectral and energy efficiency, and its robustness to angle errors, reciprocity mismatch, and multipath.
\end{abstract}

\section{Introduction}
Cell-free massive MIMO (CFmMIMO) delivers uniform service quality by distributing a large number of access points (APs). In frequency-division duplex (FDD) systems, downlink (DL) precoding requires channel state information (CSI) at the APs, whose acquisition demands both DL pilots for user equipment (UE)-side estimation and feedback---a twofold overhead that has led FDD to be widely perceived as infeasible for massive MIMO, steering most CFmMIMO research to time-division duplex (TDD) leveraging channel reciprocity. Nevertheless, FDD remains indispensable: paired bands dominate global spectrum allocations and cannot readily migrate to TDD for economic and regulatory reasons~\cite{Ref_flordelis2018massive}.

Prior work deals with the two overheads separately. For training, compressed-sensing techniques exploit the sparse angle-domain structure to reduce the pilot dimension~\cite{Ref_han2022downlink}. For feedback, angular reciprocity---the approximate invariance of the angle of departure (AoD) and the angle of arrival (AoA) across paired bands---allows DL CSI to be inferred from uplink (UL) measurements~\cite{Ref_abdallah2020efficient, Ref_kim2020downlink}; yet these schemes either limit precoding to channel statistics and send no DL pilots~\cite{Ref_abdallah2020efficient} or still feed back instantaneous path gains for coherent operation~\cite{Ref_kim2020downlink}. Other approaches shrink, rather than remove, the feedback---quantizing it~\cite{Ref_lee2024user} or acquiring the channel through selected ports that are still trained and fed back~\cite{Ref_zhang2024joint}.

This correspondence proposes a framework built from four components: (i)~\emph{sparse AP activation} (SAA) assigns each user its strongest available AP, collapsing the DL pilot dimension so that orthogonal training and coherent detection become affordable while the idle APs' circuits are switched off (Section~\ref{sec:framework}); (ii)~\emph{angle-reciprocity precoding} (ARP) steers each DL beam solely by UL-estimated angles, eliminating the need for CSI feedback completely (Section~\ref{sec:framework}); (iii)~a \emph{load-adaptive dual-mode optimizer} refines the ARP beams---closed-form signal-to-leakage-and-noise ratio (SLNR) precoding boosts the data rate at high traffic load, while semidefinite relaxation (SDR)-based power minimization meets per-user average signal-to-interference-plus-noise ratio (SINR) targets at low traffic load (Section~\ref{sec:optimization}); and (iv)~a \emph{closed-form achievable rate}, valid even under angle-estimation errors, reciprocity mismatch, and multipath (Sections~\ref{sec:framework} and~\ref{sec:numerical}).

\section{System Model}
Consider a CFmMIMO system, where \( Q \) \( N_{t} \)-antenna APs serve $K$ single-antenna UEs, for $M = QN_{t} \gg K$. Antenna elements are closely spaced, e.g., at half-wavelength spacing \( d = \lambda/2 \), so the array response toward an angle $\theta$ is captured by the steering vector $\mathbf{a}_q(\theta) = [1, e^{-j \frac{2\pi}{\lambda} d\sin \theta}, \ldots, e^{-j \frac{2\pi}{\lambda} (N_t-1)d\sin \theta}]^T$ at AP $q$.
Denote the AP and user index sets by $\mathbb{Q}=\{1,\ldots,Q\}$ and $\mathbb{K}=\{1,\ldots,K\}$. A central processing unit (CPU) connects all APs via fronthaul. The system operates in FDD over paired bands $\{f_u, f_d\}$; since the UL differs little between duplexing modes, we focus on the DL.

Let $g_{kq}\sim\mathcal{CN}(0,\beta_{kq})$, independent across links, denote the complex channel gain between AP $q$ and UE $k$, where $\beta_{kq}$ captures path loss and shadowing. For analytical tractability, we adopt a dominant-path abstraction: the array response is modeled as $\mathbf{h}_{kq}=g_{kq}\,\mathbf{a}_q(\theta_q^k)$, with $\theta_q^k$ the mean angle of the dominant path (line-of-sight or strongest reflected cluster) from AP $q$ to UE $k$.\footnote{This model fits the short serving links enabled by SAA, where a dominant path is likely, and becomes more accurate at higher carrier frequencies.} This narrowband model applies per OFDM subcarrier; since the angles are set by propagation geometry, they are common to all subcarriers and to both duplex bands, so the angle-domain quantities driving the framework remain frequency-flat even when the channel is not. On the DL frequency $f_d$, AP $q$ in general transmits $\mathbf{s}_q = \sum_{k'\in\mathbb{K}} \mathbf{w}_{q,k'}\,x_{k'}$, where $\mathbf{w}_{q,k'}\in\mathbb{C}^{N_t}$ is the precoder for UE $k'$ and $x_{k'}\sim\mathcal{CN}(0,1)$ is the unit-power data symbol. Under the SAA framework (Section~\ref{subsec:SAA}), each AP serves one user; we denote its precoder and data symbol by $\mathbf{w}_q$ and $x_q$, respectively ($\mathbf{w}_q=\mathbf{0}$ for APs left inactive by SAA). Thus, $\mathbf{w}_q$ captures both beam direction and transmit power, the latter bounded by the per-AP power budget $P_t$, i.e., $\|\mathbf{w}_q\|^2 \le P_t$. With the beam pattern of AP~$q$ defined as $\Phi_q(\theta) = \mathbf{a}_q^T(\theta)\mathbf{w}_q$, the received signal at UE $k$ is $y_k = \sum_{q=1}^{Q} g_{kq}\, \Phi_q(\theta_{q}^k)\, x_q + n_k$, where the additive noise $n_k\sim\mathcal{CN}(0,\sigma_n^2)$.

\section{Feedback-Free FDD-CFmMIMO Framework}\label{sec:framework}
The framework proceeds in three steps: (i) SAA selects the serving APs, (ii) ARP sets their beams from UL angles, and (iii) reduced-pilot training enables coherent detection.

\subsection{Sparse AP Activation}\label{subsec:SAA}

\begin{figure}[!t]
\centering
\includegraphics[width=0.38\textwidth]{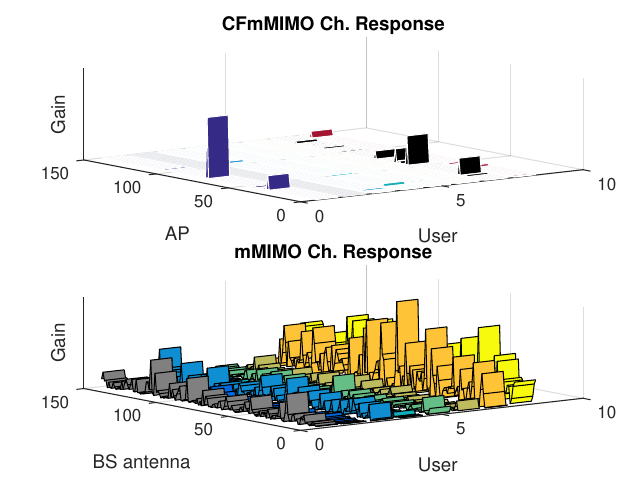}
\caption{Illustrative channel-response snapshot: collocated mMIMO versus CFmMIMO with its inherent spatial sparsity.}\label{Dia_channel}
\vspace{-2mm}
\end{figure}

In cell-free architectures,  $\beta_{kq}$ vary by orders of magnitude across APs for the same user; only a few APs are close, while most are distant, their links heavily attenuated (Fig.~\ref{Dia_channel}). This \emph{spatial sparsity}---absent in collocated massive MIMO (mMIMO), whose antennas share nearly identical large-scale statistics per user---implies that activating only a few nearby APs is efficient. We therefore exploit the slowly varying $\{\beta_{kq}\}$ to greedily associate each user with its best available AP. Initializing $\mathbb{K}_1=\mathbb{K}$ and $\mathbb{Q}_1=\mathbb{Q}$, iteration $i=1,\ldots,K$ selects the strongest remaining UE--AP pair
\begin{equation}\label{eq:saa}
(k_i,\, \mathbbmss{q}_{k_i}) = \arg\max_{(k,\,q)\,\in\,\mathbb{K}_i \times \mathbb{Q}_i} \beta_{kq},
\end{equation}
and removes it from the candidate sets, i.e., $\mathbb{K}_{i+1}=\mathbb{K}_i\setminus\{k_i\}$ and $\mathbb{Q}_{i+1}=\mathbb{Q}_i\setminus\{\mathbbmss{q}_{k_i}\}$. After $K$ iterations, every UE is paired, yielding the active set $\mathbb{Q}_s=\{\mathbbmss{q}_1,\ldots,\mathbbmss{q}_K\}$ of $K$ \emph{distinct} APs, with $\mathbbmss{q}_k$ serving UE~$k$.\footnote{When two users share the same strongest AP, the one with the weaker link takes its strongest remaining AP, which in the simulated setting affects $5\%$ of the users and changes the sum spectral efficiency by less than $0.1\%$ compared with the assignment that maximizes the sum of the gains in dB.}

\emph{Remark~1 (Generalization):} \textit{The procedure extends to associate each user with $D\ge 1$ APs by iterating the selection; while this promises macro-diversity, it also increases the number of interfering beams. Without feedback, the serving APs cannot align their phases, so the coherent gain of CFmMIMO is unattainable; in the simulated setting, $D=1$ gives the highest spectral and energy efficiency.}

\emph{Remark~2 (Why FDD Needs SAA):} \textit{Let $N_a=|\mathbb{Q}_s|$ denote the number of active APs and $\tau_c$ the coherence-block length, divided into $\tau_u$ symbols for the UL---at least $K$ of them being orthogonal UL pilots---and $\tau_d=\tau_c-\tau_u$ for the DL; the same split applies to all duplexing modes for fairness. Channel reciprocity makes TDD free of DL pilots, whereas FDD must place its DL training within $\tau_d$, so the achievable-rate prelogs for TDD, feedback-based FDD ($M$ pilots for the antenna-domain CSI its precoder requires), and SAA ($N_a$ pilots) are $\eta^{\mathrm{TDD}}=\tau_d/\tau_c$, $\eta^{\mathrm{FDD}}_{\mathrm{full}}=(\tau_d - M)/\tau_c$, and $\eta^{\mathrm{FDD}}_{\mathrm{SAA}}=(\tau_d - N_a)/\tau_c$. In TDD user-centric clustering, idle APs may be powered down opportunistically, but deactivation is optional since TDD's training cost is independent of $N_a$. In FDD, $\eta^{\mathrm{FDD}}_{\mathrm{full}}$ becomes non-positive when $M\ge\tau_d$, making full-pilot FDD infeasible. SAA restores feasibility by collapsing the pilot cost to $N_a=K\ll Q$---an $M/K$-fold reduction (e.g., $32\times$ for $Q=64$, $N_t=4$, $K=8$)---yielding a circuit/fronthaul power saving $\Delta P_{\mathrm{circ}} = (Q-N_a)(P_{\mathrm{AP}}+P_{\mathrm{fh}}-P_{\mathrm{sleep}})$, where $P_{\mathrm{AP}}$, $P_{\mathrm{fh}}$, and $P_{\mathrm{sleep}}$ denote the per-AP circuit, fronthaul, and sleep-mode power, respectively.
}

\subsection{Basic Angle-Reciprocity Precoding}

A key enabler for FDD operation is \emph{angular reciprocity}: over a typical FDD duplex gap on the order of $100$~MHz, the scattering environment remains nearly unchanged, so the AoA and AoD are almost identical across both bands. Correlating the received UL block with the orthogonal pilot of user $j$ cancels the other users' pilots and leaves, at each active AP, the $N_t$-dimensional array response of user $j$ alone, so each angle is a single-source estimate (a one-dimensional peak search, e.g., MUSIC on a rank-one covariance), one per pilot, yielding $\{\theta_q^j\}_{j\in\mathbb{K}}$ at every active AP: no joint resolvability condition on $K$ versus $N_t$ arises, and $K\le\tau_u$ orthogonal pilots is the only requirement. The design treats these estimates as exact; Section~\ref{sec:numerical} quantifies the impact of residual errors. In its basic form, ARP sets the precoding vector at AP $\mathbbmss{q}_k$ as the scaled and conjugated steering vector:
\begin{equation}\label{eq:bf_vector}
  \mathbf{w}_{\mathbbmss{q}_k} = \sqrt{\tfrac{P_t}{N_t}}\,\mathbf{a}_{\mathbbmss{q}_k}^*(\theta_{\mathbbmss{q}_k}^k),
\end{equation}
which radiates full power $\|\mathbf{w}_{\mathbbmss{q}_k}\|^2 = P_t$ and forms a main lobe toward user $k$ with coherent  gain $\Phi_{\mathbbmss{q}_k}(\theta_{\mathbbmss{q}_k}^k)=\mathbf{a}_{\mathbbmss{q}_k}^T(\theta_{\mathbbmss{q}_k}^k)\mathbf{w}_{\mathbbmss{q}_k} = \sqrt{P_t N_t}$. While this basic beam uses only the served user's angle, the full angle set drives the optimized ARP designs of Section~\ref{sec:optimization}. With SAA and ARP, the observation at UE $k$ becomes $y_k = g_{k\mathbbmss{q}_k}\sqrt{P_t N_t}\,x_{\mathbbmss{q}_k} + \sum\nolimits_{q\in\mathbb{Q}_s\setminus\{\mathbbmss{q}_k\}} g_{kq}\,\mathbf{a}_q^T(\theta_{q}^k)\,\mathbf{w}_q\,x_q + n_k$, where the sum is the inter-user interference (IUI) originating from the sidelobes of the other active APs' beams. Therefore, the instantaneous SINR for user $k$ is
\begin{equation}\label{eq:sinr}
  \gamma_k = \frac{ |g_{k\mathbbmss{q}_k}|^2\, P_t N_t}{\sum\nolimits_{q\in\mathbb{Q}_s\setminus\{\mathbbmss{q}_k\}} |g_{kq}|^2\, |\mathbf{a}_q^T(\theta_{q}^k) \mathbf{w}_q|^2 + \sigma_n^2}.
\end{equation}

\subsection{Reduced-Pilot Estimation for Coherent Detection}

Although precoder design requires only angle information at the AP, coherent detection at the UE requires knowledge of the instantaneous \emph{effective channel} $h_k \triangleq g_{k\mathbbmss{q}_k}\, \Phi_{\mathbbmss{q}_k}(\theta_{\mathbbmss{q}_k}^k)$, whose mean power is $\Omega_k \triangleq \mathbb{E}[|h_k|^2] = \beta_{k\mathbbmss{q}_k}\, |\Phi_{\mathbbmss{q}_k}(\theta_{\mathbbmss{q}_k}^k)|^2$. This formulation holds for \emph{any} precoder $\mathbf{w}_{\mathbbmss{q}_k}$; under basic ARP \eqref{eq:bf_vector}, $|\Phi_{\mathbbmss{q}_k}(\theta_{\mathbbmss{q}_k}^k)|^2 = P_t N_t$. We estimate $h_k$ through DL pilot transmission from the activated APs.
Since only $N_a = K$ APs are active, $\tau_p \geq K$ orthogonal pilots suffice.

Mutually orthogonal pilot sequences $\{\boldsymbol{\psi}_k\}_{k\in\mathbb{K}}\subset\mathbb{C}^{\tau_p}$ with $\|\boldsymbol{\psi}_k\|^2=\tau_p$, the $k$-th assigned to the $k$-th activated AP, are transmitted, giving the received pilot vector $\mathbf{y}_{k,p} = h_k\boldsymbol{\psi}_k + \sum\nolimits_{j\in\mathbb{K}\setminus\{k\}} g_{k\mathbbmss{q}_j}\Phi_{\mathbbmss{q}_j}(\theta_{\mathbbmss{q}_j}^k)\,\boldsymbol{\psi}_j + \mathbf{n}_k$, where $\mathbf{n}_k\sim\mathcal{CN}(\mathbf{0},\sigma_n^2\mathbf{I}_{\tau_p})$. Correlating with $\boldsymbol{\psi}_k$ removes all cross-AP terms exactly and, after normalization by $\sqrt{\tau_p}$, yields the sufficient statistic $y_{k,p} = \sqrt{\tau_p}\, h_k + \tilde{n}_k$ with $\tilde{n}_k\sim\mathcal{CN}(0,\sigma_n^2)$. Minimum mean-square error (MMSE) estimation then gives $\hat{h}_k = \mathbb{E}[y_{k,p}^*h_k]\,y_{k,p}/\mathbb{E}[|y_{k,p}|^2] = \sqrt{\tau_p}\,\Omega_k\, y_{k,p}/(\tau_p \Omega_k + \sigma_n^2) \sim \mathcal{CN}(0, \alpha_k)$ with estimation variance $\alpha_k = \tau_p\Omega_k^2/(\tau_p \Omega_k + \sigma_n^2)$; the estimation error $\tilde{h}_k = h_k - \hat{h}_k \sim \mathcal{CN}(0, \Omega_k - \alpha_k)$ is independent of $\hat{h}_k$.

\emph{Remark~3 (DL pilots and coherent detection):} \textit{
Channel hardening is significantly weaker in cell-free systems than in collocated arrays; relying on statistical CSI alone severely degrades rates~\cite{Ref_interdonato2019downlink}. SAA makes orthogonal DL pilots affordable, so each UE detects coherently instead. Unlike in limited-feedback schemes~\cite{Ref_kim2020downlink, Ref_lee2024user}, the estimate remains at the UE and is never fed back.
}

\subsection{Achievable Spectral and Energy Efficiency}

With the MMSE decomposition $h_k = \hat{h}_k + \tilde{h}_k$ above, the UE knows $\hat{h}_k$ and detects coherently. Conditioned on $\hat{h}_k$, the estimation error, the IUI, and the thermal noise are uncorrelated with the signal and have aggregate variance $(\Omega_k-\alpha_k) + I_k + \sigma_n^2$, where
$I_k = \sum_{q\in\mathbb{Q}_s\setminus\{\mathbbmss{q}_k\}}
    \beta_{kq}\,|\Phi_q(\theta_q^k)|^2$
captures the IUI; treating this aggregate as worst-case Gaussian noise yields the achievable-rate lower bound $R_k = \eta^{\mathrm{FDD}}_{\mathrm{SAA}}\,\mathbb{E}\bigl[\log_2\bigl(1 + \bar{\gamma}_k\,|\hat{h}_k|^2/\alpha_k\bigr)\bigr]$, with average effective SINR $\bar{\gamma}_k = \alpha_k/((\Omega_k-\alpha_k)+I_k+\sigma_n^2)$. Since $|\hat{h}_k|^2/\alpha_k$ is unit-mean exponential, the expectation admits the exact \textit{closed form}, in terms of the exponential integral,
\begin{equation}\label{eq:rate}
  R_k = \eta^{\mathrm{FDD}}_{\mathrm{SAA}}\, \log_2(e)\,
     e^{1/\bar{\gamma}_k} \int_{1/\bar{\gamma}_k}^{\infty} z^{-1}e^{-z}\,\mathrm{d}z.
\end{equation}
Expression \eqref{eq:rate} is exact for single-serving-AP schemes. Under basic ARP, $\Omega_k = P_t N_t \beta_{k\mathbbmss{q}_k}$, the mean of the numerator of \eqref{eq:sinr}.
The sum spectral efficiency (SE) is $\mathrm{C}_{\mathrm{total}}=\sum_{k=1}^{K}R_k$ (bps/Hz).

The total consumed power comprises transmit, circuit, fronthaul, and sleep-mode components~\cite{Ref_ngo2018total}, $P_{\mathrm{total}} = \frac{1}{\mu}\sum_{q\in\mathbb{Q}_s}\|\mathbf{w}_q\|^2 + N_a(P_{\mathrm{AP}} + P_{\mathrm{fh}}) + (Q - N_a)\,P_{\mathrm{sleep}}$, where $\mu\in(0,1]$ is the power-amplifier efficiency and the remaining quantities were defined in Remark~2; the energy efficiency (EE, in bit/J) is $\mathrm{EE} = B\,\mathrm{C}_{\mathrm{total}}/P_{\mathrm{total}}$, with $B$ the system bandwidth.

\section{Load-Adaptive Dual-Mode Precoding}\label{sec:optimization}
 Network traffic varies, but whatever the traffic model, the scheduler turns the demand of each user into a rate target, which corresponds to a per-user average-SINR target $\xi_k$ at the precoder. We therefore design a dual-mode ARP optimizer, with a rate-boosting mode and a power-saving mode, that adapts to this demand. For the scheduled targets $\{\xi_k\}$, the CPU first checks whether they can be met in the power-saving mode, i.e., whether the semidefinite program (SDP) \eqref{EQN:minPower_Relaxation} below is feasible. If so, \emph{SDR-based precoding} serves them with minimum power. If not, the demand exceeds the capability, and \emph{full-power SLNR precoding} boosts the sum rate by suppressing the residual IUI. The feasibility of \eqref{EQN:minPower_Relaxation} is thus the switching rule. Both modes refine the basic ARP beam \eqref{eq:bf_vector} and use only the slowly varying $\{\beta_{kq}\}$ and $\{\theta_q^k\}$, so switching costs no extra signaling.

\subsection{Rate-Boosting SLNR Precoding}
At high load, the sum SE is limited by IUI: the leakage each beam casts onto the other $K-1$ users. Each active AP thus radiates full power while refining its basic ARP beam to suppress it. Since sum-SE maximization over $\{\mathbf{w}_q\}$ is nonconvex, we adopt the SLNR criterion, which decouples across APs. Define the Hermitian positive-semidefinite matrix $\mathbf{A}_q(\theta) \triangleq \mathbf{a}_q^*(\theta)\mathbf{a}_q^T(\theta)$, so that $|\mathbf{a}_q^T(\theta)\mathbf{w}_q|^2 = \mathbf{w}_q^H\mathbf{A}_q(\theta)\mathbf{w}_q$. With the fast-fading powers averaged out ($\mathbb{E}[|g_{jq}|^2]=\beta_{jq}$), the SLNR of AP~$\mathbbmss{q}_k$ relates the power delivered to its served user to that leaked onto the other $K-1$ users, plus noise, and---absorbing the noise through the full-power constraint, $\sigma_n^2 = \mathbf{w}_{\mathbbmss{q}_k}^H(\sigma_n^2/P_t)\,\mathbf{I}_{N_t}\mathbf{w}_{\mathbbmss{q}_k}$ for $\|\mathbf{w}_{\mathbbmss{q}_k}\|^2 = P_t$---takes the form of a generalized Rayleigh quotient:
\begin{equation}\label{eq:slnr_def}
\mathrm{SLNR}_k = \frac{\beta_{k\mathbbmss{q}_k}\bigl|\mathbf{a}_{\mathbbmss{q}_k}^T(\theta_{\mathbbmss{q}_k}^k)\,\mathbf{w}_{\mathbbmss{q}_k}\bigr|^2}{\sum\nolimits_{j\ne k}\beta_{j\mathbbmss{q}_k}\bigl|\mathbf{a}_{\mathbbmss{q}_k}^T(\theta_{\mathbbmss{q}_k}^j)\,\mathbf{w}_{\mathbbmss{q}_k}\bigr|^2 + \sigma_n^2} = \frac{\mathbf{w}_{\mathbbmss{q}_k}^H\,\mathbf{B}_k\,\mathbf{w}_{\mathbbmss{q}_k}}{\mathbf{w}_{\mathbbmss{q}_k}^H\,\mathbf{C}_k\,\mathbf{w}_{\mathbbmss{q}_k}},
\end{equation}
where $\mathbf{B}_k \triangleq \beta_{k\mathbbmss{q}_k}\mathbf{A}_{\mathbbmss{q}_k}(\theta_{\mathbbmss{q}_k}^k)$ and $\mathbf{C}_k \triangleq \sum\nolimits_{j\ne k}\beta_{j\mathbbmss{q}_k}\mathbf{A}_{\mathbbmss{q}_k}(\theta_{\mathbbmss{q}_k}^j) + (\sigma_n^2/P_t)\,\mathbf{I}_{N_t} \succ \mathbf{0}$. The maximizer of \eqref{eq:slnr_def} is available in closed form: AP~$\mathbbmss{q}_k$ transmits $\mathbf{w}_{\mathbbmss{q}_k} = \sqrt{P_t}\,\mathbf{u}_k$, with $\mathbf{u}_k$ the unit-norm principal generalized eigenvector of the pair $(\mathbf{B}_k, \mathbf{C}_k)$; the achieved SLNR equals the associated largest generalized eigenvalue---a single $N_t\times N_t$ eigendecomposition per active AP, with no iterative solver involved.

\subsection{Power-Saving SDR-Based Precoding}
At low load, where the users' service requirements are fixed, the CPU instead minimizes the radiated power $\sum_q\|\mathbf{w}_q\|^2$. Since only the large-scale gains are available at the CPU, the targets $\xi_k$ are imposed on the average SINR $\tilde{\gamma}_k$, obtained from \eqref{eq:sinr} by replacing each $|g_{kq}|^2$ by its mean $\beta_{kq}$. Accordingly, the power minimization with per-user targets $\tilde{\gamma}_k\ge\xi_k$ is
\begin{equation}
\begin{aligned}
\label{EQN:sumRate_Optimization}
&\min_{\{\mathbf{w}_q\in\mathbb{C}^{N_t}\}_{q\in\mathbb{Q}_s}}
\quad \sum\nolimits_{q\in\mathbb{Q}_s} \|\mathbf{w}_q\|^2 \\
\text{s.t.} \quad &
\frac{\beta_{k\mathbbmss{q}_k}\,
|\mathbf{a}_{\mathbbmss{q}_k}^T(\theta_{\mathbbmss{q}_k}^k)
\mathbf{w}_{\mathbbmss{q}_k}|^2}
{\sum\nolimits_{q\in\mathbb{Q}_s\setminus\{\mathbbmss{q}_k\}}
\beta_{kq}\,|\mathbf{a}_q^T(\theta_q^k)\mathbf{w}_q|^2
+ \sigma_n^2}
\geq \xi_k, \,\forall k\in\mathbb{K}, \\
& \|\mathbf{w}_q\|^2 \leq P_t, \quad \forall q \in \mathbb{Q}_s,
\end{aligned}
\end{equation}
a nonconvex quadratically constrained quadratic program.

Changing the optimization variables to $\{\mathbf{W}_q \coloneqq \mathbf{w}_q\mathbf{w}_q^H\}_{q\in\mathbb{Q}_s}$---an equivalent representation if and only if $\mathbf{W}_q$ is Hermitian positive semidefinite, $\mathbf{W}_q \succeq \mathbf{0}$, with $\mathrm{rank}(\mathbf{W}_q)=1$---turns every quadratic form in \eqref{EQN:sumRate_Optimization} into a linear trace, i.e., $\left|\mathbf{a}_q^T(\theta)\mathbf{w}_q\right|^2 = \mathbf{w}_q^H \mathbf{A}_q(\theta) \mathbf{w}_q = \mathrm{tr}(\mathbf{A}_q(\theta)\mathbf{W}_q)$, where $\mathrm{tr}(\cdot)$ denotes the trace operator. The lifted problem is thus linear in $\{\mathbf{W}_q\}$ over the convex positive-semidefinite cone, its only nonconvexity being the rank-one constraints; dropping them is precisely the semidefinite relaxation~\cite{Ref_luo2010semidefinite}. Clearing each SINR fraction then yields the SDP
\begin{equation}
\begin{aligned}
\label{EQN:minPower_Relaxation}
&\min_{\{\mathbf{W}_q\}_{q\in\mathbb{Q}_s}}
\quad \sum\nolimits_{q\in\mathbb{Q}_s} \text{tr}(\mathbf{W}_q) \\
\text{s.t.} \quad &
\beta_{k\mathbbmss{q}_k}\,
\text{tr}(\mathbf{A}_{\mathbbmss{q}_k}(\theta_{\mathbbmss{q}_k}^k)
\mathbf{W}_{\mathbbmss{q}_k}) \geq \\
&\quad \xi_k \Bigl(\sum\nolimits_{q\in\mathbb{Q}_s\setminus\{\mathbbmss{q}_k\}}\!
\beta_{kq}\,\text{tr}(\mathbf{A}_q(\theta_q^k)\mathbf{W}_q)
+ \sigma_n^2\Bigr), \,\forall k\in\mathbb{K},\\
& \text{tr}(\mathbf{W}_q) \leq P_t,\;
\mathbf{W}_q \succeq \mathbf{0}, \, \forall q\in\mathbb{Q}_s,
\end{aligned}
\end{equation}
which is solvable via CVX~\cite{Ref_grant2020cvx}.

Since \eqref{EQN:minPower_Relaxation} relaxes \eqref{EQN:sumRate_Optimization}, its minimum power cannot exceed that of \eqref{EQN:sumRate_Optimization}. Here the two minima are equal, i.e., the relaxation is tight, because any feasible point $\{\mathbf{W}_q\}_{q\in\mathbb{Q}_s}$ of \eqref{EQN:minPower_Relaxation} can be turned into beams that are feasible for \eqref{EQN:sumRate_Optimization} at no more power. These beams are given by
\begin{equation}\label{eq:recovery}
\mathbf{w}_{\mathbbmss{q}_k} = \frac{\mathbf{W}_{\mathbbmss{q}_k}\,\mathbf{a}_{\mathbbmss{q}_k}^*(\theta_{\mathbbmss{q}_k}^k)}{\sqrt{\mathrm{tr}\bigl(\mathbf{A}_{\mathbbmss{q}_k}(\theta_{\mathbbmss{q}_k}^k)\mathbf{W}_{\mathbbmss{q}_k}\bigr)}},\quad k\in\mathbb{K}.
\end{equation}
They keep the desired power of every user, $|\mathbf{a}_{\mathbbmss{q}_k}^T(\theta_{\mathbbmss{q}_k}^k)\mathbf{w}_{\mathbbmss{q}_k}|^2 = \mathrm{tr}(\mathbf{A}_{\mathbbmss{q}_k}(\theta_{\mathbbmss{q}_k}^k)\mathbf{W}_{\mathbbmss{q}_k})$. They also satisfy $\mathbf{w}_q\mathbf{w}_q^H \preceq \mathbf{W}_q$ by the Cauchy--Schwarz inequality, so no interference term and no transmit power increases. Applied to the optimum of \eqref{EQN:minPower_Relaxation}, \eqref{eq:recovery} therefore yields the optimal beams of \eqref{EQN:sumRate_Optimization}, and \eqref{EQN:sumRate_Optimization} is feasible if and only if \eqref{EQN:minPower_Relaxation} is.

\subsection{Complexity Analysis}
Complexity is counted in complex multiplications. The framework comprises four stages: (i) SAA~\eqref{eq:saa} runs $K$ iterations, each searching at most $Q\times K$ large-scale fading coefficients, for $\mathcal{O}(QK^2)$ total; (ii)~angle estimation, one single-source MUSIC/ESPRIT estimate per UL pilot at each active AP, costs $\mathcal{O}(N_t^3)$ per angle, $\mathcal{O}(K^2 N_t^3)$ overall; (iii) the SLNR precoder solves one $N_t\times N_t$ generalized eigenproblem per active AP in closed form without any iterative solver---$\mathcal{O}(N_t^3)$ each, $\mathcal{O}(K N_t^3)$ total---hence the extensive evaluations of Section~\ref{sec:numerical} use SLNR precoding, with the SDP evaluated only at selected targets; (iv) the SDR-based SDP needs $\mathcal{O}(\sqrt{KN_t}\log(1/\epsilon))$ interior-point iterations ($\epsilon$: solution accuracy), each costing $\mathcal{O}(K^2 N_t^3 + K^3 N_t^2 + K^3)$ for the $K$ blocks of size $N_t$~\cite{Ref_luo2010semidefinite}. Since $N_t$ is small (e.g., $N_t=4$) and $K$ moderate, both modes remain lightweight; as all stages depend only on $\{\beta_{kq},\theta_q^k\}$, which vary on the large-scale-fading timescale (hundreds to thousands of coherence intervals), the amortized per-block cost is negligible.

\begin{figure*}[!t]
\centerline{
\subfloat[]{
\includegraphics[width=0.68\columnwidth]{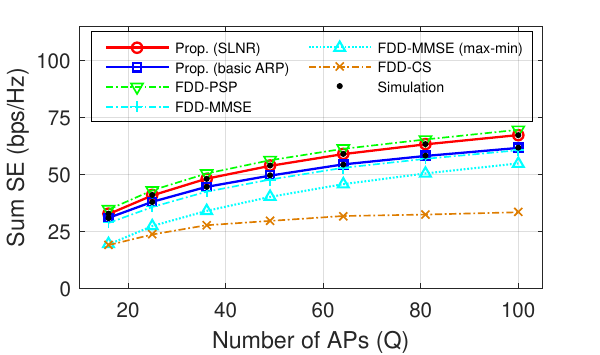}
\label{fig:se_vs_Q}
}
\hspace{-3mm}
\subfloat[]{
\includegraphics[width=0.68\columnwidth]{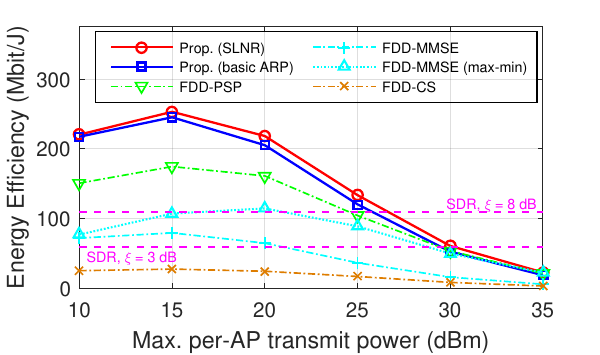}
\label{fig:ee_vs_P}
}
\hspace{-3mm}
\subfloat[]{
\includegraphics[width=0.68\columnwidth]{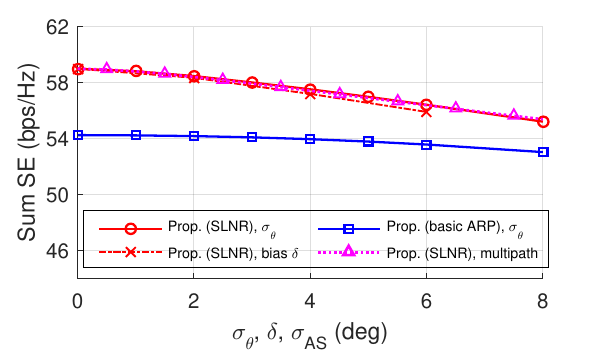}
\label{fig:robustness}
}
}
\caption{Performance comparison: (a) sum SE vs.\ $Q$, black dots marking Monte Carlo simulation of \eqref{eq:rate}; (b) EE vs.\ per-AP power budget, flat lines marking SDR-precoding EE at $\xi=3$ and $8$~dB; (c) sum SE vs.\ angle-error std $\sigma_\theta$ (solid), reciprocity bias $\delta$ (dash-dotted), and angular spread $\sigma_{\mathrm{AS}}$ of multipath with no dominant path (dotted). }

\vspace{-2mm}
\end{figure*}

\section{Numerical Results}\label{sec:numerical}

Monte Carlo simulations use the parameters of Table~\ref{tab:params}, with the APs on a square grid and the users dropped uniformly: each coherence block allocates $\tau_u=K$ UL symbols (pilots only), leaving $\tau_d=\tau_c-K=192$ DL symbols, and results are averaged over 500 independent realizations. For a fair comparison, the feedback-based baselines are evaluated by effective-channel simulation in the same way as \eqref{eq:rate}, with feedback that is error-free and costs no rate. Here $K=8$ counts the users scheduled simultaneously on one resource block ($12.5\%$ of $Q$), not the connected population.

We compare: (i)~\emph{Prop.~(SLNR)}:
SAA with the closed-form SLNR precoding;
(ii)~\emph{Prop.~(basic ARP)}: SAA with the basic ARP beam \eqref{eq:bf_vector}; (iii)~\emph{FDD-PSP}: path-selective precoding (PSP) with statistical-SLNR beams
over the $L$ strongest APs per user, with $KL$ precoded DL pilots
and $6$-bit compressed path-gain feedback ($L$ sum-SE-optimized),
adapted from~\cite{Ref_kim2020downlink}; (iv)~\emph{FDD-MMSE}: the feedback-free scheme of~\cite{Ref_abdallah2020efficient}: each user is served by the APs that hold $95\%$ of its channel power, each with the angle-domain MMSE beam, the best of its beam designs, and an equal share of the AP's power; \emph{FDD-MMSE (max-min)} uses its max-min power control instead; both are granted the same $K$ DL pilots and evaluated in the same way as \eqref{eq:rate};
(v)~\emph{FDD-CS}: compressed-sensing channel estimation from
non-orthogonal DL pilots with UL feedback of the recovered
coefficients and conjugate beamforming over each user's $S$ strongest APs,
adapted from~\cite{Ref_han2022downlink}.

\emph{1) Spectral Efficiency vs.\ Number of APs:}
In Fig.~\ref{fig:se_vs_Q}, the simulation markers coincide with the
analytical curves, confirming the correctness of our analysis~\eqref{eq:rate}. Prop.~(SLNR) achieves $67$~bps/Hz at $Q=100$ from UL angles and $K$ DL pilots, feedback-free.
Prop.~(basic ARP) follows with $62$~bps/Hz; the SLNR refinement adds
$\sim\!5$~bps/Hz by shaping the residual IUI. FDD-PSP, the strongest cooperative FDD baseline, exceeds Prop.~(SLNR) by $3$--$6\%$ ($70$~bps/Hz, narrowing with $Q$) because its fed-back gains phase-align $L$ serving APs (median $L=2$)---a margin bought with a feedback link, $KL=16$ DL pilots (twice ours), and about $14$ awake APs costing $31\%$ in EE (Fig.~\ref{fig:ee_vs_P}). Its $6$-bit quantization is near-lossless ($0.2$~bps/Hz below ideal feedback), so this margin is real. FDD-CS saturates near $33$~bps/Hz,
as the $KSN_t$ recovered coefficients fed back per coherence block are limited to $\tau_d$, which caps the support at $S=6$ APs. FDD-MMSE reaches $61$~bps/Hz, $10\%$ below Prop.~(SLNR): without feedback, the serving APs of a user cannot align their phases, so they add power, not amplitude. Its max-min power control lowers this to $55$~bps/Hz, as it lifts the weakest user at the expense of the others.

\begin{table}[t]
\centering
\caption{Simulation Parameters}
\label{tab:params}
\renewcommand{\arraystretch}{1.0}
\setlength{\tabcolsep}{3pt}
\footnotesize
\begin{tabular}{l|c||l|c}
\hline \hline
\textbf{Parameter} & \textbf{Value} & \textbf{Parameter} & \textbf{Value} \\
\hline \hline
$Q$ / $N_t$ / $K$ & 64 (default) / 4 / 8 & Coverage area & $1\!\times\!1$~km$^2$ \\
$f_d$ / FDD gap & 2~GHz / 100~MHz & Bandwidth $B$ & 20~MHz \\
Coherence $\tau_c$ & 200~sym. & Noise figure & 9~dB \\
PA efficiency $\mu$ & 0.4 & TX budget/AP $P_t$ & 200~mW \\
$P_{\mathrm{AP}}$ / $P_{\mathrm{fh}}$ & 200 / 100~mW & $P_{\mathrm{sleep}}$ & 10~mW \\
\hline
\multicolumn{4}{l}{$\beta_{kq}\text{[dB]} \!=\! -30.5 - 36.7\log_{10}(d_{kq}\text{[m]})$, shadowing $\sigma_{\text{sf}}\!=\!4$~dB} \\
\hline \hline
\end{tabular}
\end{table}

\emph{2) Energy Efficiency vs.\ Transmit Power:}
In Fig.~\ref{fig:ee_vs_P}, the proposed schemes achieve higher EE than FDD-MMSE and FDD-CS, peaking near $250$~Mbit/J at
$P_t=15$~dBm---deactivating $(Q-K)=56$ APs saves $16.24$~W of circuit
and fronthaul power. FDD-PSP peaks at $174$~Mbit/J, $31\%$ below, since its serving APs (about $14$) stay awake; FDD-MMSE peaks at $79$~Mbit/J, about a third of ours, as about $29$ of its APs stay awake to serve the $8$ users. FDD-MMSE (max-min), which radiates less, peaks at $114$~Mbit/J, under half of ours, and reaches the level of the proposed schemes only at $35$~dBm. FDD-CS, with all APs awake, remains below $28$~Mbit/J. At high power all EEs fall as the radiated power dominates; this is the SDR-precoding range, where the flat lines mark its EE at $\xi=3$ and $8$~dB (SDP
\eqref{EQN:minPower_Relaxation}).

\begin{figure}[!t]
\centering
\includegraphics[width=0.82\columnwidth]{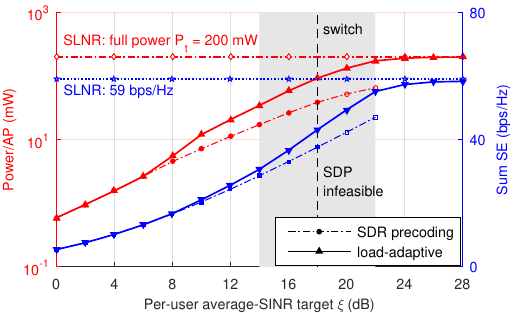}
\caption{Load-adaptive precoding over the common minimum SINR target $\xi$ ($Q=64$, $200$ realizations): radiated power per AP (left, red) and sum SE (right, blue). Dash-dotted: SDR mode, averaged over the realizations in which \eqref{EQN:minPower_Relaxation} is feasible, markers filled where this is at least half of them and hollow otherwise; the curves end where it falls below $15\%$. Solid: the load-adaptive rule, SDR where \eqref{EQN:minPower_Relaxation} is feasible and full-power SLNR otherwise. Shaded: 10th--90th percentiles of the largest feasible target; dashed: its median.}
\label{fig:qos_power}
\vspace{-2mm}
\end{figure}

\emph{3) SDR-Based Precoding and the Number of Activated APs:}
Fig.~\ref{fig:qos_power} sweeps the common target $\xi$ ($\xi_k=\xi\;\forall k$) in
\eqref{EQN:minPower_Relaxation} ($Q=64$): the SDP is feasible in all realizations up to $6$~dB, in $95\%$ at $12$~dB, $67\%$ at $18$~dB, and $20\%$ at $22$~dB; the largest feasible target has a median of $18$~dB (10th--90th percentiles $14$--$22$~dB). The load-adaptive curves leave the SDR curves where \eqref{EQN:minPower_Relaxation} turns infeasible and merge into the SLNR lines by $24$~dB. Meeting $\xi=3$~dB costs $1.2$~mW average radiated
power per active AP ($0.6\%$ of the budget) while delivering $8.9$~bps/Hz; even $\xi=12$~dB
needs only $11$~mW. Generalizing SAA to
$D$ APs per user (Remark~1) brings no SE benefit at
$Q=64$: the sum SE peaks at $D=1$ ($59.1$~bps/Hz, the
Fig.~\ref{fig:se_vs_Q} operating point, vs. $57.6$~bps/Hz at
$D=2$) and decreases monotonically, while the EE about halves ($170\to87$~Mbit/J): without feedback the serving APs'
phases cannot align, so extra activations add power, not amplitude,
while enlarging interference---cross-AP coherence in FDD
is attainable only at the feedback price quantified above.

\emph{4) Robustness to Angle Errors, Reciprocity Mismatch, and Multipath:}
Imperfect estimation is captured by
perturbing every design angle: $\hat{\theta}_q^k=\theta_q^k+\varepsilon$,
$\varepsilon\sim\mathcal{N}(0,\sigma_\theta^2)$, applied to all beams
while the channel keeps the true angles. The closed form
\eqref{eq:rate} remains exact under mismatch---the effective channel
stays conditionally Gaussian, only with a reduced beam gain. Fig.~\ref{fig:robustness} shows
the sum SE at $Q=64$: degradation is graceful, as the $N_t=4$ main lobe
is wide (half-power beamwidth $\approx 26^\circ$). Prop.~(SLNR)
loses under $1\%$ at $\sigma_\theta=2^\circ$ (the accuracy of
MUSIC/ESPRIT) and degrades by only $6\%$ at $8^\circ$.
Its SLNR gain over Prop.~(basic ARP) narrows from $4.7$ to
$2.2$~bps/Hz, since the interference nulls degrade faster than the main
lobe, but it never reverses. A deterministic reciprocity bias $\delta$
degrades the sum SE comparably to a random error of equal size ($57.2$ at
$\delta=4^\circ$ vs.\ $57.5$ at $\sigma_\theta=4^\circ$), so larger
duplex gaps follow the same curve.
Moreover, under multipath, Prop.~(SLNR) loses only $6.1\%$ in the worst case shown, Rayleigh fading with $20$ scattered paths per link and an angular spread of $\sigma_{\mathrm{AS}}=8^\circ$ (see Fig.~\ref{fig:robustness}). With a dominant path (dominant-path power ratio $\kappa=9$~dB), the loss falls to $1.3\%$.

\section{Conclusion}
This correspondence has shown that jointly exploiting spatial sparsity and angular reciprocity makes FDD CFmMIMO practical. Simulations confirmed about $10\%$ higher spectral efficiency and more than twice the peak energy efficiency of the feedback-free scheme of~\cite{Ref_abdallah2020efficient}; the only FDD baseline with higher SE requires the feedback channel and doubled training that this design eliminates. The framework is also robust: rates fall by under $1\%$ at practical angle-estimation accuracy and degrade gracefully under reciprocity bias and multipath, while the closed-form rate remains valid.

\bibliography{main}

\begin{thebibliography}{10}
\providecommand{\url}[1]{#1}
\csname url@samestyle\endcsname
\providecommand{\newblock}{\relax}
\providecommand{\bibinfo}[2]{#2}
\providecommand{\BIBentrySTDinterwordspacing}{\spaceskip=0pt\relax}
\providecommand{\BIBentryALTinterwordstretchfactor}{4}
\providecommand{\BIBentryALTinterwordspacing}{\spaceskip=\fontdimen2\font plus
\BIBentryALTinterwordstretchfactor\fontdimen3\font minus
  \fontdimen4\font\relax}
\providecommand{\BIBforeignlanguage}[2]{{\expandafter\ifx\csname l@#1\endcsname\relax
\typeout{** WARNING: IEEEtran.bst: No hyphenation pattern has been}\typeout{** loaded for the language `#1'. Using the pattern for}\typeout{** the default language instead.}\else
\language=\csname l@#1\endcsname
\fi
#2}}
\providecommand{\BIBdecl}{\relax}
\BIBdecl

\bibitem{Ref_flordelis2018massive}
J.~Flordelis, F.~Rusek, F.~Tufvesson, E.~G. Larsson, and O.~Edfors, ``Massive
  {MIMO} performance---{TDD} versus {FDD}: What do measurements say?''
  \emph{{IEEE} Trans. Wireless Commun.}, vol.~17, no.~4, pp. 2247--2261, Apr.
  2018.

\bibitem{Ref_han2022downlink}
T.~Han and D.~Zhao, ``Downlink channel estimation in {FDD} cell-free massive
  {MIMO},'' \emph{Phys. Commun.}, vol.~52, Jun. 2022, art. no. 101614.

\bibitem{Ref_abdallah2020efficient}
A.~Abdallah and M.~M. Mansour, ``Efficient angle-domain processing for
  {FDD}-based cell-free massive {MIMO} systems,'' \emph{{IEEE} Trans. Commun.},
  vol.~68, no.~4, pp. 2188--2203, Apr. 2020.

\bibitem{Ref_kim2020downlink}
S.~Kim, J.~W. Choi, and B.~Shim, ``Downlink pilot precoding and compressed
  channel feedback for {FDD}-based cell-free systems,'' \emph{{IEEE} Trans.
  Wireless Commun.}, vol.~19, no.~6, pp. 3658--3672, Jun. 2020.

\bibitem{Ref_lee2024user}
K.~Lee, J.~H. Lee, and W.~Choi, ``User-centric association and feedback bit
  allocation for {FDD} cell-free massive {MIMO},'' \emph{{IEEE} Trans. Wireless
  Commun.}, vol.~25, pp. 1891--1907, 2026.

\bibitem{Ref_zhang2024joint}
C.~Zhang, P.~Du, M.~Ding, Y.~Jing, and Y.~Huang, ``Joint port selection based
  channel acquisition for {FDD} cell-free massive {MIMO},'' \emph{{IEEE} Trans.
  Commun.}, vol.~72, no.~6, pp. 3572--3586, Jun. 2024.

\bibitem{Ref_interdonato2019downlink}
G.~Interdonato, H.~Q. Ngo, P.~Frenger, and E.~G. Larsson, ``Downlink training
  in cell-free massive {MIMO}: A blessing in disguise,'' \emph{{IEEE} Trans.
  Wireless Commun.}, vol.~18, no.~11, pp. 5153--5169, Nov. 2019.

\bibitem{Ref_ngo2018total}
H.~Q. Ngo, L.-N. Tran, T.~Q. Duong, M.~Matthaiou, and E.~G. Larsson, ``On the
  total energy efficiency of cell-free massive {MIMO},'' \emph{IEEE Trans.
  Green Commun. Netw.}, vol.~2, no.~1, pp. 25--39, Mar. 2018.

\bibitem{Ref_luo2010semidefinite}
Z.-Q. Luo, W.-K. Ma, A.~M.-C. So, Y.~Ye, and S.~Zhang, ``Semidefinite
  relaxation of quadratic optimization problems,'' \emph{{IEEE} Signal Process.
  Mag.}, vol.~27, no.~3, pp. 20--34, May 2010.

\bibitem{Ref_grant2020cvx}
M.~C. Grant and S.~P. Boyd, ``{CVX}: Matlab software for disciplined convex
  programming, version 2.2,'' \url{http://cvxr.com/cvx}, Jan. 2020.

\end{thebibliography}

\end{document}